\documentclass[10pt, a4paper]{article}

\usepackage{lrec-coling2024}
\usepackage{graphicx}
\usepackage{algorithm}
\usepackage{algorithmic}
\usepackage{multirow}
\usepackage{amsmath}
\usepackage{cleveref}
\usepackage{xspace}
\usepackage{booktabs}
\usepackage{pifont}
\usepackage{xcolor}
\usepackage[caption = false, labelformat=simple]{subfig}
\usepackage{ulem}

\crefname{table}{Table }{Tables}
\crefname{figure}{Figure }{Figures}
\crefname{section}{Section }{Sections}

\newcommand{\blockBatch}{combine-divide\xspace}
\newcommand{\model}{SEA\xspace}

\def\ie{\textit{i.e.}~}

\title{Tackling Long Code Search with Splitting, Encoding, and Aggregating}

\name{
Fan Hu$^1$\sthanks{\ \ Work done during internship at Microsoft Research Asia.} , Yanlin Wang$^2$\sthanks{\ \ The corresponding author.  Contact: Yanlin Wang (wangylin36@mail.sysu.edu.cn).} \sthanks{\ \ Work done at Microsoft Research Asia.} , Lun Du$^3$, \\ {\bf \large  Hongyu Zhang$^4$, Shi Han$^3$, Dongmei Zhang$^3$, Xirong Li$^1$} 
} 

\address{
        $^1$ Renmin University of China \\
	$^2$ School of Software Engineering, Sun Yat-sen University \\
	$^3$ Microsoft\\
	$^4$ Chongqing University\\
}

\abstract{
Code search with natural language helps us reuse existing code snippets. Thanks to the Transformer-based pretraining models, the performance of code search has been improved significantly. However, due to the quadratic complexity of multi-head self-attention, there is a limit on the input token length. For efficient training on standard GPUs like V100, existing pretrained code models, including GraphCodeBERT, CodeBERT, RoBERTa (code), take the first 256 tokens by default, which makes them unable to represent the complete information of \textit{long code} that is greater than 256 tokens. To tackle the long code problem, we propose a new baseline \model (Split, Encode and Aggregate), which splits long code into code blocks, encodes these blocks into embeddings, and aggregates them to obtain a comprehensive long code representation. With \model, we could directly use Transformer-based pretraining models to model long code without changing their internal structure and re-pretraining. We also compare \model with sparse Trasnformer methods. With GraphCodeBERT as the encoder, \model achieves an overall mean reciprocal ranking score of 0.785, which is 10.1\% higher than GraphCodeBERT on the CodeSearchNet benchmark, justifying \model as a strong baseline for long code search. 
\\ \newline  }

\begin{document}

\maketitleabstract

\section{Introduction}

A good code search technique helps developers to boost software development by searching for code snippets using natural language. Recent advancements have demonstrated the effectiveness of Transformer-based code pre-training methods, including CodeBERT \cite{20_code_bert}, CoCLR \cite{huang2021cosqa}, and GraphCodeBERT \cite{iclr_graphcodebert}, which have significantly improved code search performance through self-supervised pre-training on large-scale code corpus. 

\begin{figure}[t] 
\centering
\includegraphics[width=0.42\textwidth]{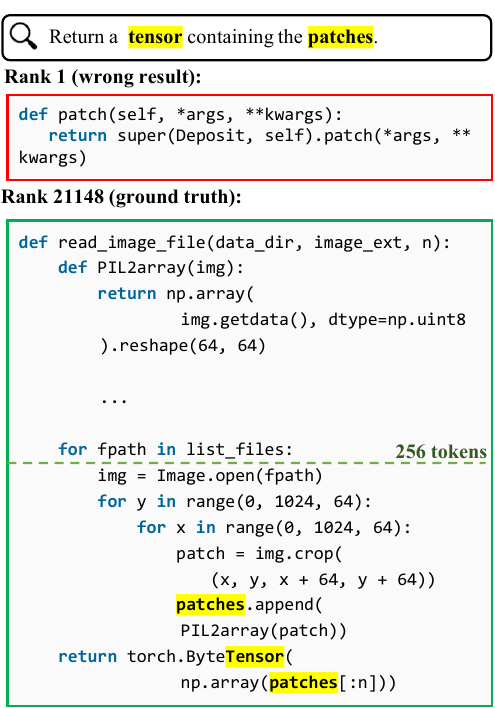} 
\caption{{Example case of GraphCodeBERT.} 
GraphCodeBERT truncates tokens beyond 256 tokens. Key tokens are highlighted in yellow.} 
\label{Figure:GraphCodeBERT_Case}
\end{figure}

However, these approaches face an inherent limitation. The computational and memory complexity of self-attention in the original Transformer grows quadratically with the input length, imposing a constraint on the input length of approximately 512 tokens. For efficient training on standard GPUs like V100, GraphCodeBERT and CodeBERT consider only the first 256 tokens of code snippets and discard any tokens beyond this limit. Nonetheless, this length restriction can lead to accuracy issues, especially for long code snippets. For instance, when examining the challenging cases of GraphCodeBERT, we found that GraphCodeBERT has low performance for some long code snippets where crucial information resides towards the end.  As illustrated in \cref{Figure:GraphCodeBERT_Case}, the keywords ``Tensor'' and ``patches'' appear after the 256-token cutoff set by GraphCodeBERT, resulting in their exclusion from consideration. Consequently, the corresponding code snippet is ranked at position 21,148.

We further conducted empirical studies on GraphCodeBERT in publicly used CodeSearchNet dataset \cite{husain2019codesearchnet}, and observed a gradual decrease in search performance as the length of the ground-truth code in the query increased (refer to \cref{tab:Empirical_study_python}). This issue is similar to the long text problem in natural language processing, for which various approaches have been proposed, including hierarchical processing \cite{zhang2019hibert}, sparse attention \cite{child2019generating, beltagy2020longformer}, and segment-level recurrence \cite{dai2019transformer}. However, directly applying these methods to long code presents two challenges. Firstly, these techniques modify the internal structure of the Transformer model, potentially rendering the existing pre-training parameters invalid. Secondly, long code differs from long text in that it is a highly structured language. Unlike a long text document that can be treated as a cohesive whole with complete semantics, the semantics of code are discontinuous, and different functions are distributed across various locations. The comparison experiments conducted in \cref{sec:compare_longformer} provide evidence supporting these concerns.

Therefore, our goal is to divide long code while preserving its semantic information. We aim to achieve this without altering the internal structure of Transformer-based pretraining models or requiring re-pretraining. To address this, we propose \model (\underline{S}plit, \underline{E}ncode, and \underline{A}ggregate) to handle long code and obtain improved code representations.

As depicted in \cref{Figure:split_and_fusion}, the process involves splitting the long code into a set of code pieces, followed by utilizing the sliding window method to generate a partially overlapping code block set. Existing code encoders are then used to obtain embeddings for each code block. Finally, these embeddings are aggregated to generate representations for the entire long code. Through extensive experiments, we have found that the proposed AST-based splitting method and attention-based aggregation method outperform other techniques for splitting and aggregation. 
Due to the varying numbers of code blocks obtained from different code snippets, parallel operation becomes challenging. To address this problem, we have designed a \blockBatch module for acceleration. It is important to note that \model is encoder-agnostic, meaning it can be used with different Transformer-based encoders. When compared to various Transformer-based encoder baselines, \model achieves a significant improvement in mean reciprocal ranking (MRR) performance, ranging from 7\% to 10\%.

The contributions can be summarized as:
\begin{sloppypar}
\begin{itemize}
    \item Empirical finding and verification of the difficulty for modeling long code in existing Transformer-based code search models.
    \item We propose a new baseline \model and explore an optimal  splitting and aggregation setting. We also design a \blockBatch module for acceleration.
    \item Through extensive experiments, we show the effectiveness of the proposed \model with different encoder baselines in six programming languages, resulting in a strong baseline for code search. Our source code and experimental data are available at: \url{https://github.com/fly-dragon211/SEA}.
    
\end{itemize}
\end{sloppypar}

\begin{table}[t]
\centering 
\setlength{\tabcolsep}{4pt}
\caption{{The code search performance (MRR) of different ground-truth code token lengths.} We set the code truncation length from 50 to 512. The highest results in each column are highlighted. Dataset: CodeSearchNet python. Model: GraphCodeBERT.}\label{tab:Empirical_study_python}
\scalebox{0.85}{
\begin{tabular}{lrrrrr}
\toprule
\multirow{2}{*}{\textbf{Token length}} & \multicolumn{5}{c}{\textbf{Code truncation length}} \\ \cline{2-6} 
 & \textbf{50} & \textbf{100} & \textbf{256} & \textbf{400} & \textbf{512} \\ \midrule
{[}0, 256) & \textbf{0.6274} & 0.6856 & 0.6909 & 0.6897 & 0.6906 \\
{[}256, 512) & 0.6239 & \textbf{0.7027} & \textbf{0.7237} & 0.7258 & 0.7265 \\
{[}512, 768) & 0.6004 & 0.6467 & 0.7168 & 0.7180 & 0.7181 \\
{[}768, 1024) & 0.6038 & 0.6315 & 0.7111 & \textbf{0.7375} & \textbf{0.7276} \\
{[}1024, 1943) & 0.6202 & 0.6573 & 0.6589 & 0.6835 & 0.6825 \\ \bottomrule
\end{tabular}
}
\end{table}

\section{Related Work}
\subsection{Code Search Methods}
Early studies~\cite{nie2016query, yang2017iecs, rosario2000latent, hill2011improving, satter2016search, lv2015codehow,van2017combining} in code search mainly applied information retrieval (IR) techniques directly, treating code search as a text matching task. Both queries and code snippets were considered plain text, and traditional text matching algorithms such as bag-of-words (BOW)~\cite{schutze2008introduction}, Jaccard~\cite{jaccard1901etude}, term frequency-inverse document frequency (TF-IDF)~\cite{robertson1976relevance}, BM25 (an improved version of TF-IDF)~\cite{robertson2009probabilistic}, and the extended boolean model~\cite{lv2015codehow} were employed. Since code length has minimal impact on modeling complexity, these methods could encode long code without truncation.

Following the introduction of the large-scale pre-training model BERT \cite{Devlin2019bert}, CodeBERT was proposed by \citet{20_code_bert}. CodeBERT is a model pre-trained on unlabeled source code and comments, which achieved impressive performance in text-based code search through fine-tuning on text-code paired datasets. \citet{huang2021cosqa} introduced CoCLR, a contrastive learning method that enhances query-code matching. \citet{sun2022code} developed a context-aware code translation technique that translates code snippets into natural language descriptions. \citet{gu2022accelerating} utilized deep hashing and code classification to accelerate code search, while \citet{chai2022cross} adapted few-shot meta-learning to code search. \citet{iclr_graphcodebert} proposed GraphCodeBERT, incorporating structure-aware pre-training tasks to improve code understanding and performance. 
Recently, \citet{hu2023revisiting} utilized a two-stage fusion code search framework that combines bi-encoders and cross-encoders to enhance performance. 
However, the computational complexity of Transformers and limited GPU memory often lead to the truncation of long code snippets.

\subsection{Neural Code Representation with Code Structure}

Recently, there have been notable advancements in neural code representation methods that leverage code structure, particularly Abstract Syntax Trees (AST), yielding impressive performance \cite{alon2020structural, sun2020treegen, bui2021treecaps, kim2021code, peng2021integrating, hellendoorn2019global, allamanis2021self, georgiev2022heat,ma2023capturing,du2023pre}. MMAN \cite{WanSSXZ0Y19} incorporates a multi-modal attention fusion layer to combine AST and Control Flow Graph (CFG) representations. ASTNN \cite{zhang2019novel} and CAST \cite{shi2021cast} segment large ASTs into sequences of smaller statement trees, encoding them into vectors by capturing the lexical and syntactical information of each statement. TBCAA \cite{chen2019capturing} employs a tree-based convolution network over API-enhanced ASTs. UniXcoder \cite{guo2022unixcoder} leverages both AST and code comments to enrich code representation. GraphCodeBERT \cite{iclr_graphcodebert} incorporates variable relations extracted from ASTs in its pre-training tasks. In our work, we specifically aim to capture and model the structural information present in long code snippets.

\subsection{Transformer for Long Text} \label{LongTrasformer}

The application of Transformer models for long text can be broadly divided into two categories: scaling up attention and enhancing the original Transformer model, and aggregation methods. The first category includes four main approaches: sparse attention \cite{child2019generating, correia2019adaptively, beltagy2020longformer, kitaev2019reformer, roy2021efficient, ainslie2020etc, jiang2020long, gunther2023jina}, recurrence \cite{dai2019transformer}, hierarchical mechanisms \cite{zhang2019hibert, gao2022long}, and compressed attention \cite{ye2019bp, guo2019star}. Sparse attention restricts each token to attend to only a subset of other tokens. Recurrence integrates recurrent neural network elements into Transformer models to extend their attention span. Hierarchical mechanisms model long input text hierarchically, from sentences to paragraphs. Compressed attention selectively compresses specific parts of the input. 

The second category, aggregation methods, involves aggregating multiple passage scores or representations for a long document. For instance, \citet{wang2019multi} proposed a multi-passage BERT model to globally normalize answer scores across all passages in the question answer task. In the context of document ranking, SMITH \cite{yang2020beyond} learns a document representation through hierarchical sentence representation aggregation. PARADE \cite{li2020parade} employs Max, CNN, Attention, and Transformer to aggregate the passage representations. \citet{tsujimura2023contextualized} uses a sliding window method to manage long input sequences in the context of medical Named Entity Recognition tasks.

However, these methods may not be entirely suitable for highly structured code. In well-designed programs, code within the same module, such as a function, is closely interconnected, while interactions between different modules are loosely coupled, adhering to the principle of high cohesion and low coupling. Conversely, long text in natural language tends to exhibit coherence. In this paper, we investigate the applicability of long text methods in the field of code search and propose a new baseline \model for long code search.

\begin{table}[t]
\centering 
\setlength{\tabcolsep}{3pt}
\caption{{The code token length statistic of CodeSearchNet evaluation set.}}\label{tab:dataset_token_statistic}

\scalebox{0.84}{
\begin{tabular}{lrrrrrrr}
\toprule
\textbf{Length} & \multicolumn{1}{l}{\textbf{Ruby}} & \multicolumn{1}{l}{\textbf{JS}} & \multicolumn{1}{l}{\textbf{Go}} & \multicolumn{1}{l}{\textbf{Py}} & \multicolumn{1}{l}{\textbf{Java}} & \multicolumn{1}{l}{\textbf{Php}} & \multicolumn{1}{l}{\textbf{Overall}} \\ \midrule
{[}0, 256) & 16\% & 10\% & 22\% & 14\% & 13\% & 13\% & 14\% \\
{[}256, 512) & 44\% & 29\% & 38\% & 30\% & 27\% & 26\% & 32\% \\
{[}512, +$\infty$) & 41\% & 62\% & 40\% & 56\% & 60\% & 61\% & 54\% \\ \bottomrule
\end{tabular}
}
\end{table}

\section{Motivation: Long Code Problem}
\subsection{Preliminaries}
Code search aims to find the most relevant code snippet $C$ from a given codebase that matches a query $Q$. 
For a current deep-learning model, we first transform query $Q$ and the code snippets $C$ to query and code tokens with the $\mathbf{tokenizer}$ such as BPE \cite{sennrich2016neural}. Then we transform the token ids of the query $Q$ and the code snippets $C$ to vector representations $\mathbf{e_q}$ and $\mathbf{e_c}$ by neural network encoders, and calculate the similarity (or distance) measures in Euclidean space such as Cosine similarity or Euclidean distance to obtain the cross-modal similarity score $s$. The calculation can be formalized as follows:

\begin{equation}
\begin{cases}
\mathbf{e_q} = \Gamma(\mathbf{tokenizer}(Q)) \\
\mathbf{e_c} = \Gamma'(\mathbf{tokenizer}(C)), C \in Codebase \\
s = sim(\mathbf{e_q}, \mathbf{e_c})
\end{cases}
\end{equation}
where $\Gamma$ and $\Gamma'$ are two well-trained neural network encoders learned from labeled paired data.

\subsection{The Long Code Problem}
To control memory and computation costs in training stage,  it is common practice to truncate long code. For example, GraphCodeBERT typically takes the first 256 code tokens by default. To investigate whether this truncation method results in information loss, we conducted token length statistics on CodeSearchNet. As shown in \cref{tab:dataset_token_statistic}, we found that snippets with a token length less than 256 accounted for only 14.1\%, while 53.5\% of code snippets exceeded the maximum encoding length of 512 tokens for Transformers. This indicates that truncation leads to information loss for snippets with a token length greater than 256.

To examine the search performance difference of GraphCodeBERT across query subsets with varying ground truth (GT) code lengths, we divided the python test subset of CodeSearchNet (CSN) into 5 distinct query sets based on different GT code token lengths. We calculated the Mean Reciprocal Rank (MRR) of GraphCodeBERT for various code truncation lengths, as shown in \cref{tab:Empirical_study_python}. Notably, we observed a downward trend in search performance as the ground-truth code token length increased (from top to bottom) for code token lengths surpassing 256 tokens, indicating that long code snippets pose challenges for GraphCodeBERT. 
Moreover, as the code truncation length extended from left to right, we observed a relatively consistent search performance when the truncation length exceeded the token length. And there emerged an upward trend in the search performance for code snippets with the token length surpassing the truncation length.
This suggests that simply truncating long code may result in the loss of valuable information.

\begin{figure*}
\centering 
    \subfloat[AST-based code splitting. 
    \label{fig:AST_split}]
    {\includegraphics[width=1.9\columnwidth]{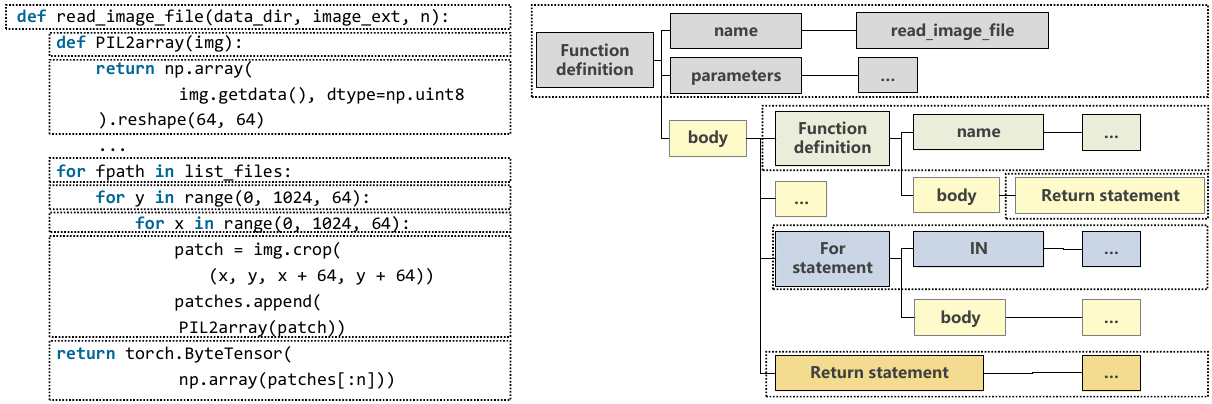}}
    
    \subfloat[Slidding window and aggregation.\label{fig:main_architecture}]
    {\includegraphics[width=1.92\columnwidth]{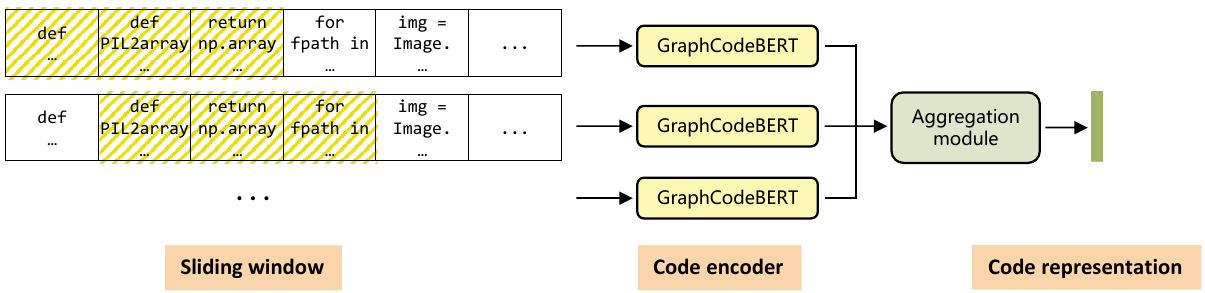}}  
    \caption{The pipeline of our proposed SEA (split, encode and aggregate) architecture. 
    }
    \vspace{-6pt}
    \label{Figure:split_and_fusion}
\end{figure*}

\section{\model}
In this section, we present a comprehensive overview of \model, encompassing the model architecture, splitting methods, aggregation techniques, and the \blockBatch method designed to accelerate inference.

\subsection{Model Architecture} \label{sec:architecture}
We introduce our \model in this section. The overall pipeline is illustrated in \cref{Figure:split_and_fusion}. Given a code snippet $C$, our objective is to derive a code representation $e_c$. To achieve this, we employ a multi-step approach. We first split the code snippet into a code piece set:
\begin{equation}
    P=\mathbf{Split}(C)=\{p_1, p_2, \dots, p_n\}.
\end{equation}
Then we use the sliding window method to obtain a partially overlapping code block set: 
\begin{equation}
    B=\mathbf{SlidingWindow}(P) = \{b_1, b_2, \dots, b_k\}.
\end{equation}
Assuming the window size is $w$ and the step is $s$, then the code block number is $k=\lfloor \frac{n-w}{s}+1 \rfloor$, where $\lfloor \cdot \rfloor$ refers to round down. Next, we utilize a code encoder, such as GraphCodeBERT, to obtain embeddings for each of the $k$ code blocks:
\begin{equation}
    e_B = \{e_{b_1}, e_{b_2}, \dots, e_{b_k}\}.
\end{equation}
Finally, an aggregation method is applied to combine the $k$ embeddings into the code representation $e_c$:
\begin{equation}
    e_c = \mathbf{Aggregation}(e_B)
\end{equation}

\subsection{Splitting Methods}\label{sec:split_methods}
To obtain the code piece set, we explore four splitting methods, namely space-based splitting, token-based splitting, line-based splitting, and AST-based splitting. Space-based splitting is simply splitting by space, resulting in splitting a string like ``def read\_image\_file'' is divided into \{`def', `read\_image\_file'\}. Similarly, token-based splitting and line-based splitting entail splitting based on tokens and lines, respectively.

An Abstract Syntax Tree (AST) is a tree representation of the syntactic structure of source code written in a programming language. Each node in the AST corresponds to a specific construct in the code, such as expressions, statements, or declarations. The hierarchical structure of ASTs reflects the syntax of programming languages, abstracting away certain syntactic details to focus on the core structure. 

For AST-based splitting, our goal is to devise a method that is both straightforward and applicable to various programming languages. Inspired by CAST \cite{shi2021cast}, we parse a source code into an Abstract Syntax Tree with tree\_sitter\footnote{\url{https://github.com/tree-sitter/py-tree-sitter}}, and visit this AST by preorder traversal. In the case of composite structures (\ie for, if, def, etc.), as depicted in \cref{fig:AST_split}, we define the set of AST nodes \{head\_block, body\}, where head\_block is responsible for splitting the header and body of nested statements such as if and While statements, while body corresponds the method declarations. When encountering a composite structure, we insert a splitting mark before and after the head\_block, effectively dividing a large AST into a sequence of non-overlapping subtrees. Subsequently, based on the AST splitting, we construct the code piece set $P$ by splitting the original code accordingly.

\subsection{Aggregation Methods} \label{sec:fusing_methods}
\textbf{Meanpooling / Maxpooling}. 
A straightforward approach to aggregate the embeddings of $k$ code blocks is to calculate the mean or maximum of their embeddings:

\begin{equation}
    e_c = \mathbf{Mean/Max}(\{e_{b_1}, e_{b_2}, \dots, e_{b_k}\}).
\end{equation}
However, a limitation of meanpooling is that each code block contributes equally to the final representation, regardless of their individual qualities. Similarly, maxpooling gives prominence to the block with the highest value. To address these limitations and enhance the aggregation process, we propose the incorporation of weighted embedding methods.

\begin{figure}[t]
\centering 
    \subfloat[One layer attention with\\ mean / max. \label{fig:fusion_one_layer}]{\includegraphics[width=0.25\textwidth]{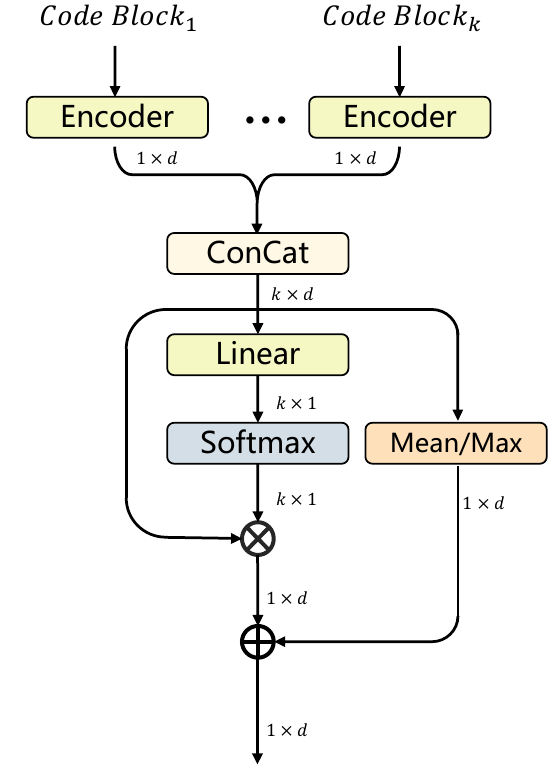}}
    \subfloat[Two layer attention with\\ mean / max. \label{fig:fusion_two_layer}]{\includegraphics[width=0.245\textwidth]{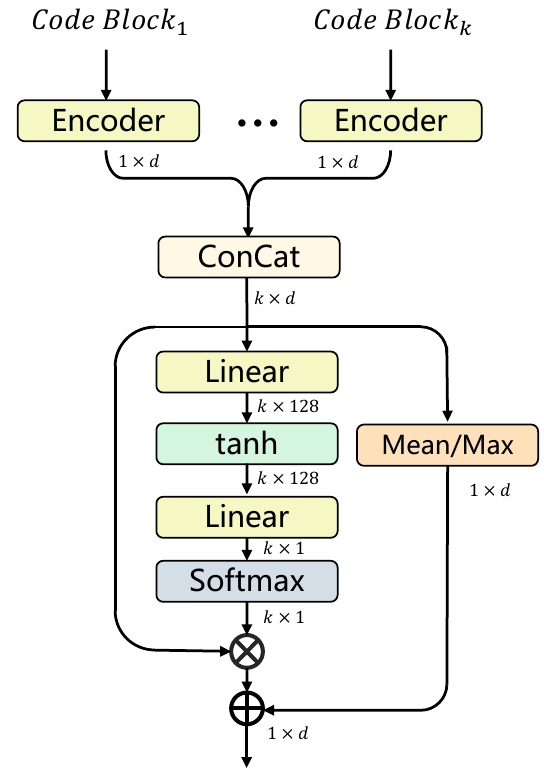}}
\caption{{The attention-based aggregation methods.} 
    }
\label{Figure:fusion_one_layer}
\end{figure}

\textbf{Attention-based aggregation}. 
Recognizing that not all code blocks hold equal importance in representing long code snippets, we introduce self-adaptive weights $\alpha$ for each block embedding during aggregation:
\begin{equation}
    e_c = \sum_i^{k}\alpha_i e_{b_i}.
\end{equation}
Inspired by attention-based Multi-Instance Learning \cite{li2021multi} and  Lightweight Attentional Feature Fusion \cite{hu2022lightweight}, we compute the weights $\{\alpha_1, \dots, \alpha_k\}$ as follows:
\begin{equation}\label{eq:light-att}
\{a_1, \ldots, a_k\} = softmax(Linear(\{e_{b_1}, \ldots, e_{b_k}\})).
\end{equation}
For one layer attention, $Linear$ refers to a fully connected layer that transforms the dimension to 1. For two layer attention, $Linear$ refers to two fully connected layers that first transform the dimension to 128 and then transform the dimension to 1. Furthermore, as illustrated in \cref{fig:fusion_one_layer} and \cref{fig:fusion_two_layer}, we explore the combination of attention with meanpooling / maxpooling methods:
\begin{equation}
    e_c = \sum_i^{k}(\alpha_i e_{b_i}) + \mathbf{Mean/Max}(\{e_{b_1}, e_{b_2}, \dots, e_{b_k}\}).
\end{equation}

For computation cost analysis, \model employs the sliding window method to significantly reduce complexity to $1/k$. The original complexity of GraphCodeBERT is given by $O(n^2 \cdot d \cdot l)$, where $n, d, l$ represent sequence length, representation dimension, and layer number, respectively. By using the sliding window method, the complexity for each window becomes $O(w^2 \cdot d \cdot l)$, where $w$ denotes the window size. Setting the step $s = w$, the total number of code blocks becomes $k = \frac{n}{w}$, leading to the window size $w = \frac{n}{k}$. Consequently, the overall complexity is simplified to:
\begin{equation}
O(k \cdot w^2 \cdot d \cdot l) = O(k \cdot {(\frac{n}{k})}^2 \cdot d \cdot l) = O(\frac{n^2}{k} \cdot d \cdot l).
\end{equation}
This remarkable reduction in complexity to $\frac{1}{k}$ allows \model to encode \textit{long code} with less memory and computation costs.

Furthermore, as shown in \cref{tab:Computation_cost}, we observe that compared to GraphCodeBERT, \model incorporating one-layer attention Aggregation introduces only $d$ additional learnable parameters. Despite this modest increase in parameter count, it plays a pivotal role in enhancing the effectiveness of the aggregation stage, as our experiments will provide the evidence in \cref{sec:exp_Optimal_conf}.

\begin{table}[t]
\centering
\setlength{\tabcolsep}{7pt}
\begin{center}
\caption{{Computation cost analysis. $n$ is the sequence length, $d$ is the representation dimension, $k$ is the code block number, $l$ is the layer number. Note that we use one layer attention for \model.}}
\label{tab:Computation_cost}
\scalebox{0.99}{
\begin{tabular}{@{}lll@{}}
\toprule
\textbf{Method} & \textbf{Parameters} & \textbf{Complexity}  \\ \midrule
GraphCodeBERT & $5d^2 \cdot l$ & $O(n^2 \cdot d \cdot l)$ \\
\model & $5d^2 \cdot l + d$ & $O(\frac{n^2}{k} \cdot d \cdot l)$
\\ \bottomrule
\end{tabular}
}
\end{center}
\end{table}

\subsection{Batch Processing} \label{sec:batch_processing}
To enhance inference efficiency on large datasets, it is necessary to devise a batch processing method capable of encoding multiple long code snippets simultaneously. As outlined in \cref{sec:architecture}, we obtain multiple code blocks from each long code snippet. However, due to the varying number of corresponding code blocks for different long code snippets, devising a general batch processing approach poses a challenge.

\begin{figure}[t]
\centering 
\includegraphics[width=0.48\textwidth]{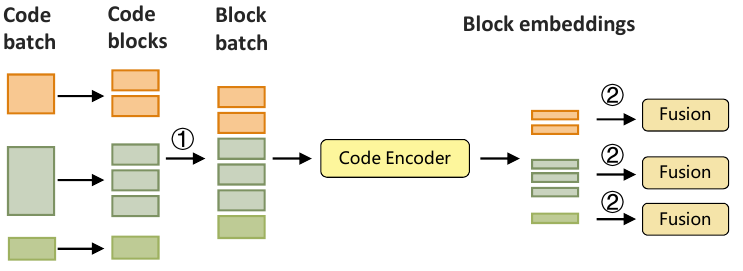}
\caption{{The batch processing \blockBatch method. } \ding{172} and \ding{173} refer to combination and division methods.}
\label{Figure:combine_divide}
\end{figure}

To address this issue, we introduce the \textbf{\blockBatch} method. As illustrated in \cref{Figure:combine_divide}, assuming a batch size of 3 (comprising three code snippets), the corresponding number of code blocks for each snippet is 2, 3, and 1, respectively. We begin by combining these six code blocks into a \textit{block batch} and establish a mapping $M$ that links the code index to the block index. Subsequently, we input this block batch into the code encoder in parallel to obtain block embeddings. Finally, leveraging the information from mapping $M$, we segregate the embeddings into three groups and input them into the aggregation module to obtain distinct code representations.

\section{Experimental Design}

\subsection{Datasets}


\begin{table*}[t]
\centering 
\caption{The search performance of different \model variants. Dataset: CodeSearchNet Ruby. } 
\label{tab:model_variation}

\setlength{\tabcolsep}{5.0pt}
\scalebox{0.80}{
\begin{tabular}{@{}l|rrrl|rrrrr@{}}
\toprule
 & \multicolumn{1}{l}{\textbf{Window}} & \multicolumn{1}{l}{\textbf{Step}} & \textbf{Splitting} & \textbf{Aggregation} & \multicolumn{1}{l}{\textbf{MRR}} & \multicolumn{1}{l}{\textbf{R@1}} & \multicolumn{1}{l}{\textbf{R@5}} & \multicolumn{1}{l}{\textbf{R@10}} & \multicolumn{1}{l}{\textbf{R@100}} \\ \midrule
GraphCodeBERT & -- & -- & -- & -- & 0.6948 & 59.3 & 82.1 & 87.3 & 96.5 \\ \midrule
\multirow{6}{*}{\model-SpaceSplitting} & 256 & 128 & Space & Maxpooling & 0.6919 & 58.5 & 82.0 & 87.2 & 95.2 \\
 & 256 & 128 & Space & Meanpooling & 0.6929 & 58.3 & 83.0 & 87.4 & 95.6 \\
  & 256 & 128 & Space & Attention (two layers) & 0.6940 & 58.7 & 83.4 & 87.1 & 94.8 \\
 & 256 & 128 & Space & Attention (two layers) + Mean & 0.7490 & 66.3 & 85.2 & 88.9 & 94.4 \\
 & 256 & 128 & Space & Attention (one layer) & 0.6989 & 59.6 & 82.2 & 86.8 & 95.0 \\
 & 256 & 128 & Space & Attention (one layer) + Mean & 0.7495 & 66.1 & 86.3 & 89.0 & 94.3 \\
 & 128 & 64 & Space & Attention (one layer) + Mean & 0.7545 & 66.2 & 87.5 & 90.2 & 95.2 \\
 & 64 & 32 & Space & Attention (one layer) + Mean & 0.7431 & 65.1 & 85.6 & 88.7 & 94.0 \\ \midrule
\multirow{3}{*}{\model-TokenSplitting} & 256 & 128 & Token & Attention (one layer) + Mean & 0.7752 & 68.4 & 89.1 & 91.9 & 96.0 \\
 & 128 & 64 & Token & Attention (one layer) + Mean & 0.7606 & 67.2 & 87.5 & 91.3 & 95.6 \\
 & 64 & 32 & Token & Attention (one layer) + Mean & 0.7352 & 62.8 & 87.2 & 90.6 & 95.0 \\ \midrule
\multirow{3}{*}{\model-LineSplitting} & 64 & 32 & Line & Attention (one layer) + Mean & 0.7635 & 67.3 & 88.2 & 91.3 & 95.6 \\
 & 32 & 16 & Line & Attention (one layer) + Mean & 0.7537 & 66.1 & 87.2 & 90.3 & 95.2 \\
 & 16 & 8 & Line & Attention (one layer) + Mean & 0.7498 & 65.5 & 86.9 & 90.3 & 95.0 \\ \midrule
\multirow{3}{*}{\model-ASTSplitting} & 64 & 32 & AST & Attention (one layer) + Mean & 0.7539 & 65.7 & \textbf{91.4} & \textbf{95.0} & \textbf{97.6} \\
 & 32 & 16 & AST & Attention (one layer) + Mean & \textbf{0.7762} & \textbf{68.8} & 89.1 & 92.0 & 96.4 \\
 & 16 & 8 & AST & Attention (one layer) + Mean & 0.7744 & 68.8 & 88.7 & 91.4 & 96.3 \\ \bottomrule
\end{tabular}
}
\end{table*}

We conduct experiments on the widely used CodeSearchNet \cite{husain2019codesearchnet} dataset, comprising six programming languages, \ie, Ruby, JavaScript, Go, Python, Java, and PHP. 
Following the approach in \cite{iclr_graphcodebert}, we apply filtering to eliminate low-quality queries and expand the retrieval set to encompass the entire code corpus.

\subsection{Evaluation Metrics} 
In our evaluation, we use two popular automatic criteria: MRR (Mean Reciprocal Ranking) and R@k (top-k accuracy, k=1, 5, 10, 100). They are commonly used for in previous code search studies~\cite{lv2015codehow, GU2018Deep, Sachdev2018Retrieval,husain2019codesearchnet,20_code_bert,huang2021cosqa,iclr_graphcodebert}.  In addition, we report the number of parameter and inference time as the efficiency measure.



\subsection{Experimental Settings} 

Our baseline is GraphCodeBERT. The parameters of code and natural language encoders are initialized by GraphCodeBERT. For training, we randomly select 6 code blocks from the divided code blocks of one long code. The training batch size is 32.  For evaluation, we use all divided code blocks of one long code. The evaluated batch size is 256. All experiments are conducted on a machine with Intel Xeon E5-2698v4 2.2Ghz 20-Core CPU and two Tesla V100 32GB GPUs.

\section{Experimental Results}\label{sec:exp_results}

\subsection{The Optimal \model Configuration} \label{sec:exp_Optimal_conf}
To identify the optimal configuration for \model, we conducted experiments by varying our architecture using different code splitting methods and aggregation methods, while measuring the resulting changes in search performance. Given that the CodeSearchNet Ruby dataset is relatively small, we focused on conducting experiments on the ruby subset, and we present the results in \cref{tab:model_variation}.

In \cref{tab:model_variation} rows SpaceSplitting, we experimented with various aggregation methods as described in \cref{sec:fusing_methods}. Our findings showed that using any single aggregation method in isolation did not yield significant performance improvements compared to the GraphCodeBERT Baseline. However, upon fusing the attention method with meanpooling, we observed substantial performance enhancement. Specifically, the Attention (one layer) + Mean aggregation method improved MRR and R@1 by 7.9\% and 11.5\%, respectively. Consequently, for subsequent experiments, we opted to use the Attention (one layer) + Mean aggregation method.

In \cref{tab:model_variation} rows SpaceSplitting, TokenSplitting, LineSplitting, ASTSplitting, we explored different code split methods, as detailed in \cref{sec:split_methods}. For space and token-based splitting methods, we set the window size from 64 to 256 due to the finer granularity of division. Conversely, for line and AST-based split methods, we set the window size from 16 to 64. Notably, we observed that the AST-based split method displayed outstanding performance, achieving the highest MRR and R@1 with a window size of 32. As a result, in subsequent experiments, \model refers to \model-ASTSplitting with a window size of 32, step size of 16 and the Attention (one layer) + Mean aggregation method.

\begin{table*}[t]
\centering
\setlength{\tabcolsep}{10pt}
\begin{center}
\caption{Comparison with sparse Transformers. The notation (G) indicates that the model is initialized with GraphCodeBERT parameters. The code inference time is determined by randomly selecting 1000 codes and calculating the average inference time. We repeat each time calculating experiment three times and report the mean and standard deviation. Dataset: CodeSearchNet Ruby. \model outperforms other models significantly ($p < 0.01$).}
\label{tab:Compare_longformer}
\scalebox{0.85}{
\begin{tabular}{@{}lrrrrrrr@{}}
\toprule
\textbf{Model} & \multicolumn{1}{l}{\textbf{\#Param.}} & \multicolumn{1}{l}{\textbf{Token Length}} & \multicolumn{1}{l}{\textbf{Inference Time}} & \multicolumn{1}{l}{\textbf{MRR}} & \multicolumn{1}{l}{\textbf{R@1}} & \multicolumn{1}{l}{\textbf{R@5}} & \multicolumn{1}{l}{\textbf{R@10}} \\
\midrule
GraphCodeBERT & 124.6M & 256 & 6.3 $\pm$ 0.3ms & 0.6948 & 59.3 & 82.1 & 87.3 \\ \midrule
BIGBIRD & 127.5M & 1024 & 20.1 $\pm$ 0.2ms & 0.2952 & 19.2 & 39.8 & 51.1 \\
BIGBIRD (G) & 127.5M & 1024 & 19.8 $\pm$ 0.0ms & 0.6121 & 50.8 & 74.2 & 80.7 \\
Longformer & 148.7M & 1024 & 33.7 $\pm$ 0.2ms & 0.5128 & 39.9 & 65.3 & 72.4 \\
Longformer (G) & 148.7M & 1024 & 33.7 $\pm$ 0.1ms & 0.6595 & 55.1 & 79.4 & 84.0 \\
LongCoder & 149.6M & 1024 & 68.6 $\pm$ 0.2ms & 0.4718 & 35.8 & 61.1 & 67.8 \\ \midrule
SEA & 124.6M & - & 7.2 $\pm$ 0.5ms & \textbf{0.7762} & \textbf{68.8} & \textbf{89.1} & \textbf{92.0} \\
- w/o \blockBatch & 124.6M & - & 24.3 $\pm$ 2.4ms & \textbf{0.7762} & \textbf{68.8} & \textbf{89.1} & \textbf{92.0} \\ \bottomrule
\end{tabular}
}
\end{center}
\end{table*}

\subsection{Comparison with Three Sparse Transformers} \label{sec:compare_longformer}

In this section, we conduct a comparison between \model and three sparse Transformers, BIGBIRD \cite{zaheer2020big}, Longformer \cite{beltagy2020longformer}, and LongCoder \cite{GuoXD0M23LongCoder}. BIGBIRD and Longformer are two well-known long document-oriented Transformers.  LongCoder employs a sliding window mechanism to handle long code input for code completion.
Specifically, we leverage the  bigbird-roberta-base\footnote{https://huggingface.co/google/bigbird-roberta-base}, longformer-base-4096\footnote{https://huggingface.co/allenai/longformer-base-4096} and longcoder-base\footnote{https://huggingface.co/microsoft/longcoder-base} models, with a token length of 1024.  
Due to BIGBIRD and Longformer not being pretrained on the code dataset, we also conducted experiments to initialize BIGBIRD and Longformer with the parameters of GraphCodeBERT. The results are presented in \cref{tab:Compare_longformer}. Comparing the results before and after initializing BIGBIRD and Longformer with the parameters of GraphCodeBERT, we found that MRR results improved from 0.2952 and 0.5016 to 0.6121 and 0.6595, respectively. We attribute this performance gap to the need for re-pretraining models that were originally pretrained on natural language datasets. 
We observed that LongCoder's MRR was 0.4718, which represents a significant decrease compared to GraphCodeBERT, suggesting that LongCoder may be primarily suited for Code Completion tasks. 
We also conducted t-tests between our \model and other baselines, and the results demonstrate that \model significantly outperforms all sparse Transformer baselines ($p < 0.01$), highlighting its superior performance in the domain of code search.

In terms of model parameters and search efficiency, \model stands out as it boasts a lower parameter count and shorter inference time compared to BIGBIRD, Longformer and LongCoder. It's worth noting that \model's parameter count is closely aligned with that of GraphCodeBERT, differing only by the addition of a single attention layer. However, this minor change results in a significant boost in search performance.
We also present experimental results without employing the \blockBatch method in \cref{tab:Compare_longformer}. We observed that  while the search performance stays stable, the inference time increases by more than threefold. It highlights the considerable improvement in inference time brought about by the \blockBatch method, thereby confirming its effectiveness in accelerating the model's inference process.

\begin{table*}[t]
\centering 
\setlength{\tabcolsep}{3.0pt}
\caption{{The MRR on six languages of the CodeSearchNet dataset. \model here refers to \model-ASTSplitting with window size 32 and step 16. \model+RoBERTa refers to \model with RoBERTa as the code encoder. \model outperforms baselines significantly ($p < 0.01$).}
}\label{tab:compare_sota}
\scalebox{0.75}{
\begin{tabular}{@{}llllllll@{}}
\toprule
\textbf{Model / Method} & \textbf{Ruby} & \textbf{Javascript} & \textbf{Go} & \textbf{Python} & \textbf{Java} & \textbf{Php} & \textbf{Overall} \\ \midrule
RoBERTa & 0.587 & 0.517 & 0.850 & 0.587 & 0.599 & 0.560 & 0.617 \\
UniXcoder & 0.586 & 0.603 & 0.881 & 0.695 & 0.687 & 0.644 & 0.683 \\
CodeBERT & 0.679 & 0.620 & 0.882 & 0.672 & 0.676 & 0.628 & 0.693 \\
GraphCodeBERT & 0.703 & 0.644 & 0.897 & 0.692 & 0.691 & 0.649 & 0.713 \\ \midrule
\model+RoBERTa & 0.651 (10.9\%$\uparrow$) & 0.593 (14.6\%$\uparrow$) & 0.879 (3.5\%$\uparrow$) & 0.633 (7.9\%$\uparrow$) & 0.666 (11.1\%$\uparrow$) & 0.647 (15.6\%$\uparrow$) & 0.678 (10.0\%$\uparrow$) \\
\model+UniXcoder & 0.648 (10.7\%$\uparrow$) & 0.692 (14.8\%$\uparrow$) & 0.896 (1.8\%$\uparrow$) & 0.707 (1.7\%$\uparrow$) & 0.739 (7.5\%$\uparrow$) & 0.712 (10.5\%$\uparrow$) & 0.732 (7.3\%$\uparrow$) \\
\model+CodeBERT & 0.742 (9.3\%$\uparrow$) & 0.696 (12.3\%$\uparrow$) & 0.905 (2.6\%$\uparrow$) & 0.714 (6.2\%$\uparrow$) & 0.732 (8.3\%$\uparrow$) & 0.711 (13.2\%$\uparrow$) & 0.750 (8.3\%$\uparrow$) \\
\model+GraphCodeBERT & \textbf{0.776 (10.4\%$\uparrow$)} & \textbf{0.742 (15.2\%$\uparrow$)} & \textbf{0.921 (2.7\%$\uparrow$)} & \textbf{0.754 (8.9\%$\uparrow$)} & \textbf{0.768 (11.1\%$\uparrow$)} & \textbf{0.748 (15.3\%$\uparrow$)} & \textbf{0.785 (10.1\%$\uparrow$)}
 \\ \bottomrule
\end{tabular}
}
\end{table*}

\subsection{\model Performance on Varied Code Lengths}

\begin{figure}[t] 
\centering
\includegraphics[width=0.48\textwidth]{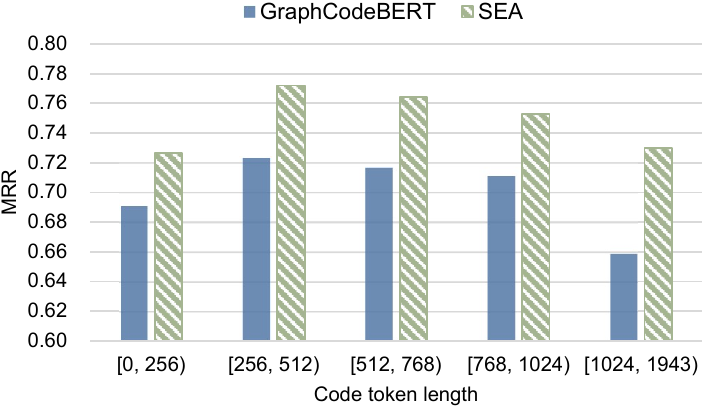}
\caption{{The performance comparison between GraphCodeBERT and \model in different ground-truth code token lengths. Compare to GraphCodeBERT, \model achieves significantly ($p < 0.01$) better performance for different code token lengths.}} 
\label{Figure:pillar_graph_ourModel}
\end{figure}

To explore the improvement of the proposed \model for code snippets with varying lengths, we present the search performance comparison between the baseline method GraphCodeBERT and \model under different ground-truth code token lengths. The results are depicted in \cref{Figure:pillar_graph_ourModel}.

Notably, the retrieval performance of each query subset exhibits noticeable enhancements, particularly for long code retrieval results. We attribute this improvement to two crucial factors. Firstly, the aggregation module of \model adaptively captures and incorporates information from diverse segments of the long code, leading to a more comprehensive and informative code representation. Secondly, the code splitting method employed by \model can be viewed as a form of data augmentation, providing additional context and variation that aids in strengthening the code representation. In summary, \model yields a more robust code representation, significantly enhancing the overall retrieval performance.

\subsection{Baseline Comparison Across Multiple Programming Languages}
To ensure a fair and reproducible comparison, we carefully selected pretraining-based baselines that meet the following three criteria: 1) The source code is publicly available;  2) The overall model is adaptable to all the six programming languages on the CodeSearchNet dataset; 3) The paper is peer-reviewed if it is published as a research paper. Consequently, we select four deep end-to-end approaches: \textbf{RoBERTa }~\cite{liu2019roberta}, \textbf{UniXcoder}~\cite{guo2022unixcoder}, \textbf{CodeBERT}~\cite{20_code_bert}, and \textbf{GraphCodeBERT}~\cite{iclr_graphcodebert}. 

In \cref{tab:compare_sota}, we present the MRR results, demonstrating that \model outperforms all methods across all six programming languages. Notably, this conclusion remains consistent for the recall metric and another variant of \model, the results of which can be found in our replication package. These findings reinforce the superiority of \model as compared to the pretraining-based baselines across diverse programming languages.

\section{Conclusion}
In this paper, we address the challenge of effectively modeling \textit{long code} for code search. We introduce \model, an effective approach that yields improved code representations for long code snippets. Despite its simplicity, our experimental results show the remarkable effectiveness and efficiency of \model. 
We believe this work opens up new possibilities for code search.

\section{Ethical Statement}
Future extensions and applications arising from our work should be mindful of the environmental impact of training large-scale models. They should actively avoid its potential misuse by searching malicious intent. However, it is unlikely that the model in its current form would lead to such an impact in the near future. Our model also has the potential for making a positive impact in areas such as code search, long code understanding and code representation.

\section{Acknowledgments}
The work described in this paper is partially supported by CCF-Huawei Populus Grove Fund CCF-HuaweiSE202301.

\section{Bibliographical References}\label{sec:reference}

\bibliographystyle{lrec-coling2024-natbib}
\bibliography{lrec-coling2024-example}

\begin{thebibliography}{60}
\expandafter\ifx\csname natexlab\endcsname\relax\def\natexlab#1{#1}\fi

\bibitem[{Ainslie et~al.(2020)Ainslie, Onta{\~{n}}{\'{o}}n, Alberti, Cvicek, Fisher, Pham, Ravula, Sanghai, Wang, and Yang}]{ainslie2020etc}
Joshua Ainslie, Santiago Onta{\~{n}}{\'{o}}n, Chris Alberti, Vaclav Cvicek, Zachary Fisher, Philip Pham, Anirudh Ravula, Sumit Sanghai, Qifan Wang, and Li~Yang. 2020.
\newblock {ETC:} encoding long and structured inputs in transformers.
\newblock In \emph{{EMNLP}}.

\bibitem[{Allamanis et~al.(2021)Allamanis, Jackson{-}Flux, and Brockschmidt}]{allamanis2021self}
Miltiadis Allamanis, Henry Jackson{-}Flux, and Marc Brockschmidt. 2021.
\newblock Self-supervised bug detection and repair.
\newblock In \emph{NeurIPS}.

\bibitem[{Alon et~al.(2020)Alon, Sadaka, Levy, and Yahav}]{alon2020structural}
Uri Alon, Roy Sadaka, Omer Levy, and Eran Yahav. 2020.
\newblock Structural language models of code.
\newblock In \emph{ICML}.

\bibitem[{Beltagy et~al.(2020)Beltagy, Peters, and Cohan}]{beltagy2020longformer}
Iz~Beltagy, Matthew~E Peters, and Arman Cohan. 2020.
\newblock Longformer: The long-document transformer.
\newblock \emph{{arXiv}}.

\bibitem[{Bui et~al.(2021)Bui, Yu, and Jiang}]{bui2021treecaps}
Nghi~DQ Bui, Yijun Yu, and Lingxiao Jiang. 2021.
\newblock Treecaps: Tree-based capsule networks for source code processing.
\newblock In \emph{AAAI}.

\bibitem[{Chai et~al.(2022)Chai, Zhang, Shen, and Gu}]{chai2022cross}
Yitian Chai, Hongyu Zhang, Beijun Shen, and Xiaodong Gu. 2022.
\newblock Cross-domain deep code search with few-shot meta learning.
\newblock \emph{arXiv}.

\bibitem[{Chen et~al.(2019)Chen, Ye, and Zhang}]{chen2019capturing}
Long Chen, Wei Ye, and Shikun Zhang. 2019.
\newblock Capturing source code semantics via tree-based convolution over api-enhanced ast.
\newblock In \emph{CF}.

\bibitem[{Child et~al.(2019)Child, Gray, Radford, and Sutskever}]{child2019generating}
Rewon Child, Scott Gray, Alec Radford, and Ilya Sutskever. 2019.
\newblock Generating long sequences with sparse transformers.
\newblock \emph{{arXiv}}.

\bibitem[{Correia et~al.(2019)Correia, Niculae, and Martins}]{correia2019adaptively}
Gon{\c{c}}alo~M Correia, Vlad Niculae, and Andr{\'e}~FT Martins. 2019.
\newblock Adaptively sparse transformers.
\newblock In \emph{{EMNLP-IJCNLP}}.

\bibitem[{Dai et~al.(2019)Dai, Yang, Yang, Carbonell, Le, and Salakhutdinov}]{dai2019transformer}
Zihang Dai, Zhilin Yang, Yiming Yang, Jaime~G Carbonell, Quoc Le, and Ruslan Salakhutdinov. 2019.
\newblock Transformer-xl: Attentive language models beyond a fixed-length context.
\newblock In \emph{{ACL}}.

\bibitem[{Devlin et~al.(2019)Devlin, Chang, Lee, and Toutanova}]{Devlin2019bert}
Jacob Devlin, Ming{-}Wei Chang, Kenton Lee, and Kristina Toutanova. 2019.
\newblock {BERT:} pre-training of deep bidirectional transformers for language understanding.
\newblock In \emph{NAACL-HLT}.

\bibitem[{Du and Yu(2023)}]{du2023pre}
Yali Du and Zhongxing Yu. 2023.
\newblock Pre-training code representation with semantic flow graph for effective bug localization.
\newblock In \emph{FSE/ESEC}.

\bibitem[{Feng et~al.(2020)Feng, Guo, Tang, Duan, Feng, Gong, Shou, Qin, Liu, Jiang, and Zhou}]{20_code_bert}
Zhangyin Feng, Daya Guo, Duyu Tang, Nan Duan, Xiaocheng Feng, Ming Gong, Linjun Shou, Bing Qin, Ting Liu, Daxin Jiang, and Ming Zhou. 2020.
\newblock Codebert: {A} pre-trained model for programming and natural languages.
\newblock In \emph{{EMNLP}}.

\bibitem[{Gao and Callan(2022)}]{gao2022long}
Luyu Gao and Jamie Callan. 2022.
\newblock Long document re-ranking with modular re-ranker.
\newblock In \emph{SIGIR}.

\bibitem[{Georgiev et~al.(2022)Georgiev, Brockschmidt, and Allamanis}]{georgiev2022heat}
Dobrik Georgiev, Marc Brockschmidt, and Miltiadis Allamanis. 2022.
\newblock Heat: Hyperedge attention networks.
\newblock \emph{arXiv}.

\bibitem[{Gu et~al.(2022)Gu, Wang, Du, Zhang, Han, Zhang, and Lyu}]{gu2022accelerating}
Wenchao Gu, Yanlin Wang, Lun Du, Hongyu Zhang, Shi Han, Dongmei Zhang, and Michael Lyu. 2022.
\newblock Accelerating code search with deep hashing and code classification.
\newblock In \emph{ACL}.

\bibitem[{Gu et~al.(2018)Gu, Zhang, and Kim}]{GU2018Deep}
Xiaodong Gu, Hongyu Zhang, and Sunghun Kim. 2018.
\newblock Deep code search.
\newblock In \emph{ICSE}.

\bibitem[{G{\"u}nther et~al.(2023)G{\"u}nther, Ong, Mohr, Abdessalem, Abel, Akram, Guzman, Mastrapas, Sturua, Wang et~al.}]{gunther2023jina}
Michael G{\"u}nther, Jackmin Ong, Isabelle Mohr, Alaeddine Abdessalem, Tanguy Abel, Mohammad~Kalim Akram, Susana Guzman, Georgios Mastrapas, Saba Sturua, Bo~Wang, et~al. 2023.
\newblock Jina embeddings 2: 8192-token general-purpose text embeddings for long documents.
\newblock \emph{arXiv}.

\bibitem[{Guo et~al.(2022)Guo, Lu, Duan, Wang, Zhou, and Yin}]{guo2022unixcoder}
Daya Guo, Shuai Lu, Nan Duan, Yanlin Wang, Ming Zhou, and Jian Yin. 2022.
\newblock Unixcoder: Unified cross-modal pre-training for code representation.
\newblock In \emph{ACL}.

\bibitem[{Guo et~al.(2021)Guo, Ren, Lu, Feng, Tang, Liu, Zhou, Duan, Svyatkovskiy, Fu, Tufano, Deng, Clement, Drain, Sundaresan, Yin, Jiang, and Zhou}]{iclr_graphcodebert}
Daya Guo, Shuo Ren, Shuai Lu, Zhangyin Feng, Duyu Tang, Shujie Liu, Long Zhou, Nan Duan, Alexey Svyatkovskiy, Shengyu Fu, Michele Tufano, Shao~Kun Deng, Colin~B. Clement, Dawn Drain, Neel Sundaresan, Jian Yin, Daxin Jiang, and Ming Zhou. 2021.
\newblock Graphcodebert: Pre-training code representations with data flow.
\newblock In \emph{{ICLR}}.

\bibitem[{Guo et~al.(2023)Guo, Xu, Duan, Yin, and McAuley}]{GuoXD0M23LongCoder}
Daya Guo, Canwen Xu, Nan Duan, Jian Yin, and Julian~J. McAuley. 2023.
\newblock Longcoder: {A} long-range pre-trained language model for code completion.
\newblock In \emph{ICML}, Proceedings of Machine Learning Research. {PMLR}.

\bibitem[{Guo et~al.(2019)Guo, Qiu, Liu, Shao, Xue, and Zhang}]{guo2019star}
Qipeng Guo, Xipeng Qiu, Pengfei Liu, Yunfan Shao, Xiangyang Xue, and Zheng Zhang. 2019.
\newblock Star-transformer.
\newblock In \emph{{NAACL-HLT}}. Association for Computational Linguistics.

\bibitem[{Hellendoorn et~al.(2019)Hellendoorn, Sutton, Singh, Maniatis, and Bieber}]{hellendoorn2019global}
Vincent~J Hellendoorn, Charles Sutton, Rishabh Singh, Petros Maniatis, and David Bieber. 2019.
\newblock Global relational models of source code.
\newblock In \emph{ICLR}.

\bibitem[{Hill et~al.(2011)Hill, Pollock, and Vijay-Shanker}]{hill2011improving}
Emily Hill, Lori Pollock, and K~Vijay-Shanker. 2011.
\newblock Improving source code search with natural language phrasal representations of method signatures.
\newblock In \emph{{ASE}}. IEEE.

\bibitem[{Hu et~al.(2022)Hu, Chen, Wang, Zhou, Dong, and Li}]{hu2022lightweight}
Fan Hu, Aozhu Chen, Ziyue Wang, Fangming Zhou, Jianfeng Dong, and Xirong Li. 2022.
\newblock Lightweight attentional feature fusion: A new baseline for text-to-video retrieval.
\newblock In \emph{ECCV}. Springer.

\bibitem[{Hu et~al.(2023)Hu, Wang, Du, Li, Zhang, Han, and Zhang}]{hu2023revisiting}
Fan Hu, Yanlin Wang, Lun Du, Xirong Li, Hongyu Zhang, Shi Han, and Dongmei Zhang. 2023.
\newblock Revisiting code search in a two-stage paradigm.
\newblock In \emph{WSDM}.

\bibitem[{Huang et~al.(2021)Huang, Tang, Shou, Gong, Xu, Jiang, Zhou, and Duan}]{huang2021cosqa}
Junjie Huang, Duyu Tang, Linjun Shou, Ming Gong, Ke~Xu, Daxin Jiang, Ming Zhou, and Nan Duan. 2021.
\newblock Cosqa: 20,000+ web queries for code search and question answering.
\newblock In \emph{{ACL}}.

\bibitem[{Husain et~al.(2019)Husain, Wu, Gazit, Allamanis, and Brockschmidt}]{husain2019codesearchnet}
Hamel Husain, Ho-Hsiang Wu, Tiferet Gazit, Miltiadis Allamanis, and Marc Brockschmidt. 2019.
\newblock Codesearchnet challenge: Evaluating the state of semantic code search.
\newblock \emph{arXiv}.

\bibitem[{Jaccard(1901)}]{jaccard1901etude}
Paul Jaccard. 1901.
\newblock {\'E}tude comparative de la distribution florale dans une portion des alpes et des jura.
\newblock \emph{Bull Soc Vaudoise Sci Nat}, pages 547--579.

\bibitem[{Jiang et~al.(2020)Jiang, Xiong, Lee, and Wang}]{jiang2020long}
Jyun-Yu Jiang, Chenyan Xiong, Chia-Jung Lee, and Wei Wang. 2020.
\newblock Long document ranking with query-directed sparse transformer.
\newblock In \emph{EMNLP Findings}.

\bibitem[{Kim et~al.(2021)Kim, Zhao, Tian, and Chandra}]{kim2021code}
Seohyun Kim, Jinman Zhao, Yuchi Tian, and Satish Chandra. 2021.
\newblock Code prediction by feeding trees to transformers.
\newblock In \emph{ICSE}.

\bibitem[{Kitaev et~al.(2019)Kitaev, Kaiser, and Levskaya}]{kitaev2019reformer}
Nikita Kitaev, Lukasz Kaiser, and Anselm Levskaya. 2019.
\newblock Reformer: The efficient transformer.
\newblock In \emph{{ICLR}}.

\bibitem[{Li et~al.(2020)Li, Yates, MacAvaney, He, and Sun}]{li2020parade}
Canjia Li, Andrew Yates, Sean MacAvaney, Ben He, and Yingfei Sun. 2020.
\newblock Parade: Passage representation aggregation for document reranking.
\newblock \emph{ACM Transactions on Information Systems}.

\bibitem[{Li et~al.(2021)Li, Zhou, Wang, Lin, Zhao, Ding, Yu, and Chen}]{li2021multi}
Xirong Li, Yang Zhou, Jie Wang, Hailan Lin, Jianchun Zhao, Dayong Ding, Weihong Yu, and Youxin Chen. 2021.
\newblock Multi-modal multi-instance learning for retinal disease recognition.
\newblock In \emph{ACMMM}.

\bibitem[{Liu et~al.(2019)Liu, Ott, Goyal, Du, Joshi, Chen, Levy, Lewis, Zettlemoyer, and Stoyanov}]{liu2019roberta}
Yinhan Liu, Myle Ott, Naman Goyal, Jingfei Du, Mandar Joshi, Danqi Chen, Omer Levy, Mike Lewis, Luke Zettlemoyer, and Veselin Stoyanov. 2019.
\newblock Roberta: A robustly optimized bert pretraining approach.
\newblock \emph{arXiv}.

\bibitem[{Lv et~al.(2015)Lv, Zhang, Lou, Wang, Zhang, and Zhao}]{lv2015codehow}
Fei Lv, Hongyu Zhang, Jian-guang Lou, Shaowei Wang, Dongmei Zhang, and Jianjun Zhao. 2015.
\newblock Codehow: Effective code search based on api understanding and extended boolean model (e).
\newblock In \emph{{ASE}}.

\bibitem[{Ma et~al.(2023)Ma, Du, and Li}]{ma2023capturing}
Y~Ma, Yali Du, and Ming Li. 2023.
\newblock Capturing the long-distance dependency in the control flow graph via structural-guided attention for bug localization.
\newblock In \emph{IJCAI}.

\bibitem[{Nie et~al.(2016)Nie, Jiang, Ren, Sun, and Li}]{nie2016query}
Liming Nie, He~Jiang, Zhilei Ren, Zeyi Sun, and Xiaochen Li. 2016.
\newblock Query expansion based on crowd knowledge for code search.
\newblock \emph{IEEE Transactions on Services Computing}, pages 771--783.

\bibitem[{Peng et~al.(2021)Peng, Li, Wang, Zhao, and Jin}]{peng2021integrating}
Han Peng, Ge~Li, Wenhan Wang, Yunfei Zhao, and Zhi Jin. 2021.
\newblock Integrating tree path in transformer for code representation.
\newblock In \emph{NeurIPS}.

\bibitem[{Robertson and Zaragoza(2009)}]{robertson2009probabilistic}
Stephen Robertson and Hugo Zaragoza. 2009.
\newblock \emph{The probabilistic relevance framework: BM25 and beyond}.
\newblock Now Publishers Inc.

\bibitem[{Robertson and Jones(1976)}]{robertson1976relevance}
Stephen~E Robertson and K~Sparck Jones. 1976.
\newblock Relevance weighting of search terms.
\newblock \emph{Journal of the American Society for Information science}, pages 129--146.

\bibitem[{Rosario(2000)}]{rosario2000latent}
Barbara Rosario. 2000.
\newblock Latent semantic indexing: An overview.
\newblock \emph{Techn. rep. INFOSYS}, pages 1--16.

\bibitem[{Roy et~al.(2021)Roy, Saffar, Vaswani, and Grangier}]{roy2021efficient}
Aurko Roy, Mohammad Saffar, Ashish Vaswani, and David Grangier. 2021.
\newblock Efficient content-based sparse attention with routing transformers.
\newblock \emph{Transactions of the Association for Computational Linguistics}, 9:53--68.

\bibitem[{Sachdev et~al.(2018)Sachdev, Li, Luan, Kim, Sen, and Chandra}]{Sachdev2018Retrieval}
Saksham Sachdev, Hongyu Li, Sifei Luan, Seohyun Kim, Koushik Sen, and Satish Chandra. 2018.
\newblock Retrieval on source code: a neural code search.
\newblock In \emph{{MAPL}}.

\bibitem[{Satter and Sakib(2016)}]{satter2016search}
Abdus Satter and Kazi Sakib. 2016.
\newblock A search log mining based query expansion technique to improve effectiveness in code search.
\newblock In \emph{{ICCIT}}, pages 586--591. IEEE.

\bibitem[{Sch{\"u}tze et~al.(2008)Sch{\"u}tze, Manning, and Raghavan}]{schutze2008introduction}
Hinrich Sch{\"u}tze, Christopher~D Manning, and Prabhakar Raghavan. 2008.
\newblock \emph{Introduction to information retrieval}, volume~39.
\newblock Cambridge University Press Cambridge.

\bibitem[{Sennrich et~al.(2016)Sennrich, Haddow, and Birch}]{sennrich2016neural}
Rico Sennrich, Barry Haddow, and Alexandra Birch. 2016.
\newblock Neural machine translation of rare words with subword units.
\newblock In \emph{{ACL}}.

\bibitem[{Shi et~al.(2021)Shi, Wang, Du, Zhang, Han, Zhang, and Sun}]{shi2021cast}
Ensheng Shi, Yanlin Wang, Lun Du, Hongyu Zhang, Shi Han, Dongmei Zhang, and Hongbin Sun. 2021.
\newblock Cast: Enhancing code summarization with hierarchical splitting and reconstruction of abstract syntax trees.
\newblock In \emph{{EMNLP}}.

\bibitem[{Sun et~al.(2022)Sun, Fang, Chen, Tao, Han, and Zhang}]{sun2022code}
Weisong Sun, Chunrong Fang, Yuchen Chen, Guanhong Tao, Tingxu Han, and Quanjun Zhang. 2022.
\newblock Code search based on context-aware code translation.
\newblock \emph{arXiv}.

\bibitem[{Sun et~al.(2020)Sun, Zhu, Xiong, Sun, Mou, and Zhang}]{sun2020treegen}
Zeyu Sun, Qihao Zhu, Yingfei Xiong, Yican Sun, Lili Mou, and Lu~Zhang. 2020.
\newblock Treegen: A tree-based transformer architecture for code generation.
\newblock In \emph{AAAI}.

\bibitem[{Tsujimura et~al.(2023)Tsujimura, Yamada, Ida, Miwa, and Sasaki}]{tsujimura2023contextualized}
Tomoki Tsujimura, Koshi Yamada, Ryuki Ida, Makoto Miwa, and Yutaka Sasaki. 2023.
\newblock Contextualized medication event extraction with striding ner and multi-turn qa.
\newblock \emph{Journal of Biomedical Informatics}, page 104416.

\bibitem[{Van~Nguyen et~al.(2017)Van~Nguyen, Nguyen, Phan, Nguyen, and Nguyen}]{van2017combining}
Thanh Van~Nguyen, Anh~Tuan Nguyen, Hung~Dang Phan, Trong~Duc Nguyen, and Tien~N Nguyen. 2017.
\newblock Combining word2vec with revised vector space model for better code retrieval.
\newblock In \emph{ICSE-C}. IEEE.

\bibitem[{Wan et~al.(2019)Wan, Shu, Sui, Xu, Zhao, Wu, and Yu}]{WanSSXZ0Y19}
Yao Wan, Jingdong Shu, Yulei Sui, Guandong Xu, Zhou Zhao, Jian Wu, and Philip~S. Yu. 2019.
\newblock Multi-modal attention network learning for semantic source code retrieval.
\newblock In \emph{{ASE}}.

\bibitem[{Wang et~al.(2019)Wang, Ng, Ma, Nallapati, and Xiang}]{wang2019multi}
Zhiguo Wang, Patrick Ng, Xiaofei Ma, Ramesh Nallapati, and Bing Xiang. 2019.
\newblock Multi-passage bert: A globally normalized bert model for open-domain question answering.
\newblock In \emph{EMNLP}.

\bibitem[{Yang et~al.(2020)Yang, Zhang, Li, Bendersky, and Najork}]{yang2020beyond}
Liu Yang, Mingyang Zhang, Cheng Li, Michael Bendersky, and Marc Najork. 2020.
\newblock Beyond 512 tokens: Siamese multi-depth transformer-based hierarchical encoder for long-form document matching.
\newblock In \emph{CIKM}.

\bibitem[{Yang and Huang(2017)}]{yang2017iecs}
Yangrui Yang and Qing Huang. 2017.
\newblock Iecs: Intent-enforced code search via extended boolean model.
\newblock \emph{Journal of Intelligent \& Fuzzy Systems}, pages 2565--2576.

\bibitem[{Ye et~al.(2019)Ye, Guo, Gan, Qiu, and Zhang}]{ye2019bp}
Zihao Ye, Qipeng Guo, Quan Gan, Xipeng Qiu, and Zheng Zhang. 2019.
\newblock Bp-transformer: Modelling long-range context via binary partitioning.
\newblock \emph{arXiv}.

\bibitem[{Zaheer et~al.(2020)Zaheer, Guruganesh, Dubey, Ainslie, Alberti, Onta{\~{n}}{\'{o}}n, Pham, Ravula, Wang, Yang, and Ahmed}]{zaheer2020big}
Manzil Zaheer, Guru Guruganesh, Kumar~Avinava Dubey, Joshua Ainslie, Chris Alberti, Santiago Onta{\~{n}}{\'{o}}n, Philip Pham, Anirudh Ravula, Qifan Wang, Li~Yang, and Amr Ahmed. 2020.
\newblock Big bird: Transformers for longer sequences.
\newblock In \emph{NeurIPS}.

\bibitem[{Zhang et~al.(2019{\natexlab{a}})Zhang, Wang, Zhang, Sun, Wang, and Liu}]{zhang2019novel}
Jian Zhang, Xu~Wang, Hongyu Zhang, Hailong Sun, Kaixuan Wang, and Xudong Liu. 2019{\natexlab{a}}.
\newblock A novel neural source code representation based on abstract syntax tree.
\newblock In \emph{ICSE}.

\bibitem[{Zhang et~al.(2019{\natexlab{b}})Zhang, Wei, and Zhou}]{zhang2019hibert}
Xingxing Zhang, Furu Wei, and Ming Zhou. 2019{\natexlab{b}}.
\newblock Hibert: Document level pre-training of hierarchical bidirectional transformers for document summarization.
\newblock In \emph{{ACL}}.

\end{thebibliography}

\end{document}